\documentclass[
prd,
twocolumn,
nofootinbib,
preprintnumbers,
eqsecnum,
superscriptaddress,
showpacs
]{revtex4}

\usepackage{graphicx}
\usepackage{amsfonts}
\usepackage{amsmath}
\usepackage{amssymb}
\usepackage{amsxtra}
\usepackage{latexsym}
\usepackage{color}
\usepackage{colordvi}
\usepackage{xcolor}
\usepackage{epsfig}
\usepackage{array}
\usepackage{pifont}
\usepackage{braket}
\newcommand{\alinea}{\hspace*{\parindent}}

\def\krto{ {\,\,\lower .8ex\hbox {$\longrightarrow \atop k \rightarrow 0$}\,\,}}

\def\alinea{\hspace{\parindent}}
\def\bea{\begin{eqnarray} }
\def\beq{\begin{eqnarray} }

\def\eea{\end{eqnarray}}
\def\eeq{\end{eqnarray}}

\begin{document} %\date{\today}
%\date{}
%%%%%%%%%%%%%%%%%%%%%%%%%%%%%%%%%%%%%%%%%%%%%%
%% title page 
%%%%%%%%%%%%%%%%%%%%%%%%%%%%%%%%%%%%%%%%%%%%%%

\title{Comment on ``Lattice Gluon and Ghost Propagators, and the Strong Coupling in Pure $SU(3)$
Yang-Mills Theory:  Finite Lattice Spacing and Volume Effects"}

\author{Ph.~Boucaud} 
\affiliation{ Laboratoire de Physique Th\'eorique (UMR8627), CNRS, Univ. Paris-Sud, Universit\'e Paris-Saclay, 91405 Orsay, France}
\author{F.~De Soto}
\affiliation{Dpto. Sistemas F\'isicos, Qu\'imicos y Naturales, 
Univ. Pablo de Olavide, 41013 Sevilla, Spain}
\author{J.~Rodr\'{\i}guez-Quintero}
\affiliation{Department of Integrated Sciences;  
University of Huelva, E-21071 Huelva; Spain.}
\affiliation{CAFPE, Universidad de Granada, E-18071 Granada, Spain}
\author{S.~Zafeiropoulos}
\affiliation{Department of Physics, College of William and Mary, Williamsburg, VA 23187-8795, USA}
\affiliation{Jefferson Laboratory, 12000 Jefferson Avenue, Newport News, VA 23606, USA}

%\author{Ph.~Boucaud$^{1}$, F.~De Soto$^{2}$, J.~Rodr\'{\i}guez-Quintero$^{3,4}$, S.~Zafeiropoulos$^{5,6}$}

\begin{abstract}

The authors of ref.~\cite{Duarte:2016iko} reported about a careful analysis of the impact of lattice artifacts on the $SU(3)$ gauge-field propagators. In particular, they found that the low-momentum behavior of the renormalized propagators depends on the lattice bare coupling and interpreted this fact as the result of it being affected by finite lattice spacing artifacts. We do not share this interpretation and present here a different and more suitable explanation for these results.

\end{abstract}


\maketitle

%\begin{flushright}
%%DAMTP-2011-nnn\\
%LPT-Orsay 11-74\\
%UHU-FT/11-29 \\
%%LPSC-11-nnn \\
%IRFU-11-136
%\end{flushright}
%%
%\vspace*{-1cm}
%%
%\begin{figure}[h]
%%  \begin{center}
%    \includegraphics[width=25mm]{figs/ETMC_rund.pdf}
%%  \end{center}
%\end{figure}

%\vfill
%\newpage

%%%%%%%%%%%%%%%%%%%%%%%%%%%%%%%%%%%%%%%%%%%%%%%%%%%%%%%%%
%% end of title page
%%%%%%%%%%%%%%%%%%%%%%%%%%%%%%%%%%%%%%%%%%%%%%%%%%%%%%%%%

%%%%%%%%%%%%%%%%%%%%%%%%%%%%%%%%%%%%%%%%%%%%%%%%%%%%%%%%%
%% body of the paper
%%%%%%%%%%%%%%%%%%%%%%%%%%%%%%%%%%%%%%%%%%%%%%%%%%%%%%%%%

\section{Introduction}
%\alinea
\vspace*{-0.1cm}

The understanding of the IR dynamics of QCD has been very much boosted in the past years by the endeavors in obtaining a very detailed picture for the fundamental Green's functions of the theory in both lattice~\cite{Cucchieri:2006tf,Cucchieri:2008qm,Cucchieri:2007md,Cucchieri:2010xr,Bogolubsky:2009dc,Oliveira:2009eh,
Ayala:2012pb,Duarte:2016iko} and continuum QCD~\cite{Aguilar:2008xm,Boucaud:2008ky,Fischer:2008uz,RodriguezQuintero:2010wy,
Pennington:2011xs,Maris:2003vk,Aguilar:2004sw,Boucaud:2005ce,Fischer:2006ub,Kondo:2006ih,Binosi:2007pi,Binosi:2008qk,
Boucaud:2007hy,Dudal:2007cw,Dudal:2008sp,Kondo:2011ab,Szczepaniak:2001rg,Szczepaniak:2003ve,Epple:2007ut,
Szczepaniak:2010fe,Watson:2010cn,Watson:2011kv}. Namely, a consensus has been reached about both the fact that the gluon propagator takes a non-zero finite value at vanishing momentum (corresponding to a dynamical generation of an effective gluon mass~\cite{Cornwall:1981zr,Bernard:1982my,Donoghue:1983fy,Philipsen:2001ip}) and the fact that the ghost propagator behaves essentially as its tree-level expression dictates. These findings have recently contributed, for instance, to establish a striking connection between gauge and matter sectors in defining an interaction kernel for a symmetry-preserving truncation of Schwinger-Dyson equations (SDEs) able to reproduce the observable properties of hadrons~\cite{Binosi:2014aea}; as well as to the construction of a process-independent strong running coupling which agrees very well with the Bjorken sum-rule effective charge~\cite{Binosi:2016nme}. 

Very recently, the authors of \cite{Duarte:2016iko} have performed a thorough study of the effect of lattice artifacts on pure Yang-Mills $SU(3)$ gluon and ghost propagators in Landau gauge, as a result of which they claimed that they both depend on the lattice spacing, $a$, in the infrared domain, while finite volume effects appear to be very mild when lattice volumes are larger than ($6.5$ fm)$^4$, in physical units. Specifically, the authors concluded that the zero-momentum gluon propagator dropped roughly by a factor of 10\% when the lattice spacing increases from $0.06$ fm ($\beta$=6.3) up to $0.18$ fm ($\beta$=5.7). This appeared to be, in our view, wrongly attributed to a discretization artifact. Indeed, these artifacts taking place at the length scale $a$ can hardly be felt by gluon modes with characteristic wavelengths of $1/p \gg a$, corresponding to deep infrared momenta. Furthermore, one should expect for them, controlled by powers of $a p$, not to be stronger at low infrared than at large UV momenta. Our intention here is to propose an alternative explanation, other than the one based on discretization artifacts, which might account for the findings described in \cite{Duarte:2016iko}. Our proposed interpretation can be confirmed by a further scrutiny of the data published therein, although we will preliminary check it here with some gluon propagator lattice data that we have recently obtained, and applied for different purposes, in ref.~\cite{Boucaud:2017obn}.

\section{Lattice scale deviations}
\alinea
\vspace*{-0.1cm}

Let us focus on the Landau-gauge gluon propagator, defined as
%----
\begin{equation}
D^{ab}_{\mu\nu}(p) =  \langle A_\mu^a(p) A_\nu^b(-p) \rangle =  
\delta^{ab} \left( \delta_{\mu \nu} - \frac{p_\mu p_\nu}{p^2} \right) D(p^2)   
\end{equation}
%----
where $A_\mu^a$ is the gauge field in momentum space, latin (greek) indices correspond to color (Lorentz) degrees of freedom, $\langle \cdot \rangle$ expresses the integration over the gauge fields, which is replaced by the average over gauge field configurations in lattice QCD,  and $D(p^2)$ is the so-called gluon propagator which, as explained in \cite{Duarte:2016iko}, is to be renormalized on the lattice by applying the MOM prescription,
%----
\begin{equation}
\left. D_R(p^2,\zeta^2) \right|_{p^2=\zeta^2} =  Z_3^{-1}(\zeta^2) D(\zeta^2) = \frac 1 {\zeta^2} \ ;
\end{equation}
%----
where $\zeta^2$ is the renormalization point, fixed at 4 GeV in ref.~\cite{Duarte:2016iko}. The details of the computation of the gluon propagator on the lattice can be found in the literature, for instance in some previous works of the authors of \cite{Duarte:2016iko}, as \cite{Silva:2004bv}, or in previous works of some of us as~\cite{Ayala:2012pb}.

\begin{figure}[t]
\begin{center}
\begin{tabular}{c}
\includegraphics[width=0.9\linewidth]{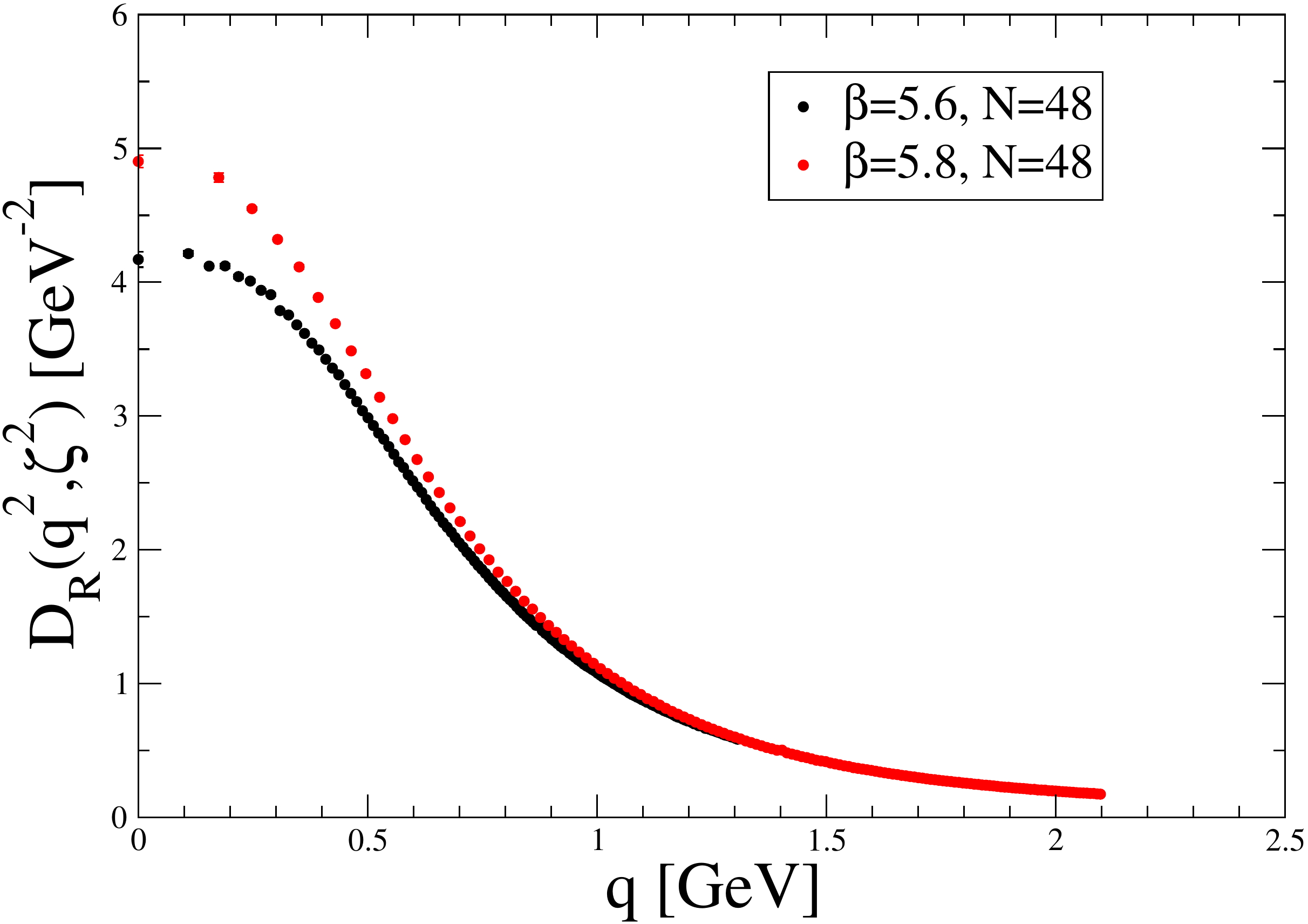} 
\\
\includegraphics[width=0.9\linewidth]{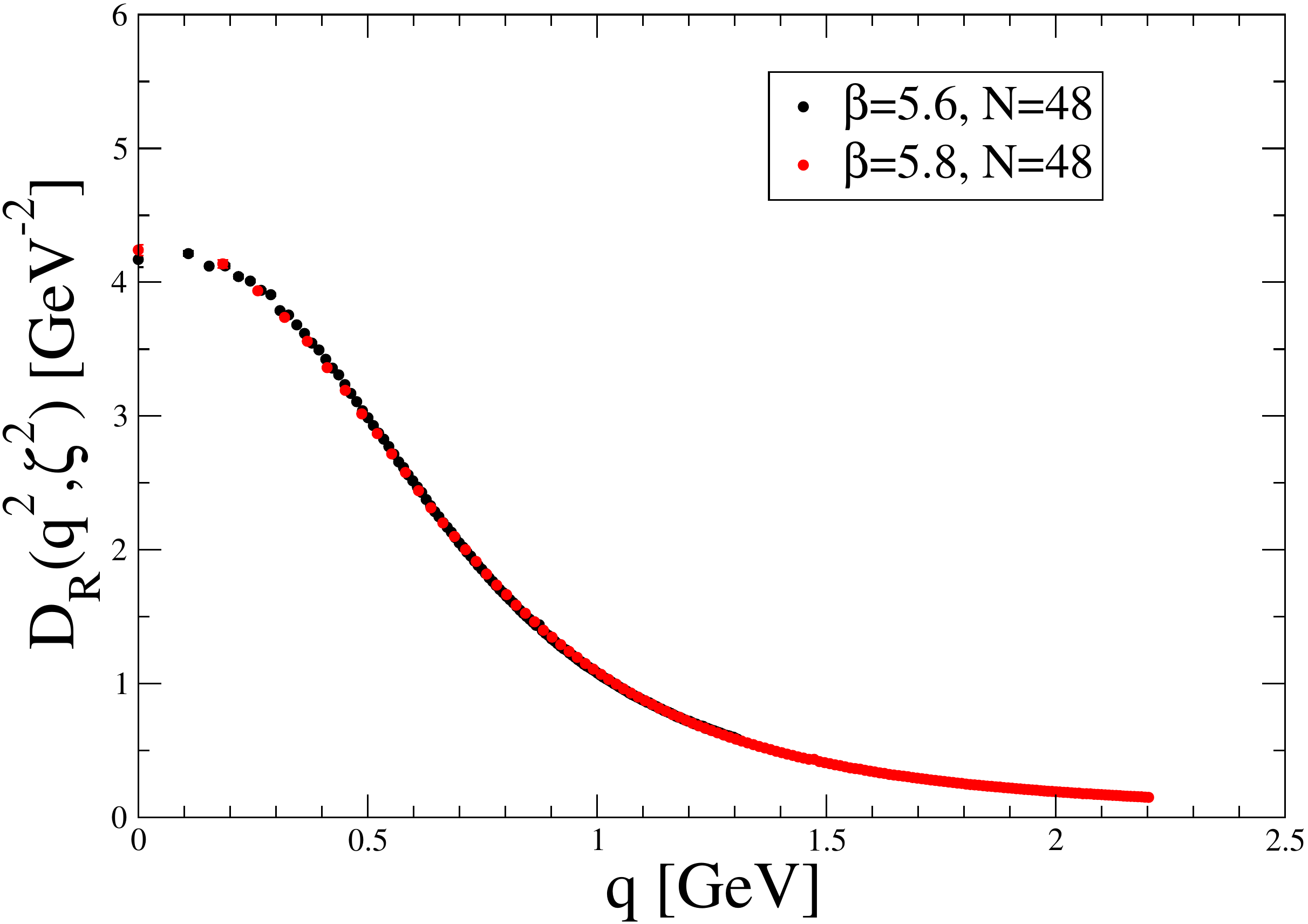} 
\end{tabular}
\vspace*{-0.3cm}
\caption{Upper panel.- Lattice gluon propagator results for the set-ups given in Tab.~\ref{tab:setup}. Lower panel.- The same gluon propagator results after applying to the data at $\beta$=5.8 the ``{\it recalibration}" described in the text through Eqs.~(\ref{eq:recal1},\ref{eq:recal2}), with $\delta$=-0.05 for the deviation parameter.}
\label{fig:prop}
\end{center}
\end{figure}

\begin{table}[h]
\begin{center}
\begin{tabular}{||cccc||}
\hline
\hline
$\beta$ & $N$ & $a$ [fm] & confs \\ 
\hline
5.6 & 48 & 0.236 & 1920 \\
5.8 & 48 & 0.147 & 960 \\
\hline
\hline
\end{tabular}
\end{center}
\vspace*{-0.3cm}
\caption{Lattice set-ups specifying the bare lattice coupling $\beta=6/g_0^2$, the number of lattice sites in any of the directions, $N$, the lattice spacing, $a$, and the number of gauge-field configurations exploited. The lattice scales has been taken from~\cite{Guagnelli:1998ud}.}
\label{tab:setup}
\end{table}

In a very recent lattice analysis of the three-gluon vertex and running coupling~\cite{Boucaud:2017obn}, we have also computed the gluon propagator for different lattice bare couplings. In particular, we obtained the results displayed in Fig.~\ref{fig:prop}, for $\beta$=5.6 and $\beta$=5.8 from quenched simulations with the Wilson action in 48$^4$ lattices. Details of the lattice set-ups can be found in Tab.~\ref{tab:setup}. The statistical errors have been estimated by applying the jackknife method. The propagators are displayed as a function of the lattice momenta $p_\mu=2\pi/(N a) n_\mu$, with $n_\mu=0,1, \dots N/4$, instead of the tree-level improved $\widehat{p}_\mu=2/a\sin{(a p_\mu/2)}.$ We have applied the $H(4)$-extrapolation~\cite{Becirevic:1999uc}, which has been proven as a very efficient prescription to cure the data from the hypercubic artifacts~\cite{Becirevic:1999uc,Becirevic:1999hj,deSoto:2007ht}. In addition, we have also employed such a kinematical cut that $a p \le \pi/2$, thus lessening the impact of any remaining discretization artifact. As a consequence of this, the largest accessible momentum for the simulation at $\beta$=5.6 is not much above the momentum, $\zeta=1.3$ GeV, which we take here for the renormalization point. Indeed, imposing the renormalization condition at $\zeta=4$ GeV, for which $a\zeta \sim 1.5 \pi$ at $\beta$=5.6 and $\sim \pi$ at $\beta$=5.8, might imply to incorporate sizable discretization artifacts and, as the propagators are thus required to take there the same value, $1/\zeta^2$, propagate these artifacts down to low IR momenta. 

The latter is a possible source, partially at least, for the lattice spacing effect reported in \cite{Duarte:2016iko}. However, our propagators displayed in the upper panel of Fig.~\ref{fig:prop}, renormalized at $\zeta=1.3$ GeV, show the same effect: the data obtained with a larger value of the lattice spacing (lower $\beta$) appear to deviate upwards when the momentum decreases. Alternatively, we claim that this striking feature cannot be a discretization artifact but the consequence of a systematic uncertainty in the lattice scale setting. Indeed, if one admits a small deviation in the lattice scale, $a^{(\delta)}=a (1+\delta)$, the ``{\it recalibrated}" gluon propagator would read as
%----
\beq\label{eq:recal1}
D^{(\delta)}(p^2) = (1+\delta)^2 D((1+\delta)^2p^2) \ , 
\eeq
and, after renormalization at $p^2=\zeta^2$, 
%----
\beq\label{eq:recal2}
D_R^{(\delta)}(p^2,\zeta^2) = \frac{D((1+\delta)^2p^2)}{\zeta^2 D((1+\delta)^2\zeta^2)} \ ;
\eeq
%----
where $D$ stands for the bare lattice propagator obtained with the lattice spacing $a$. Therefore, the systematic deviation in the scale setting expressed by $\delta$ would result in a non-trivial transformation of the data that might well account for the low-momentum discrepancies shown by the upper panel of Fig.~\ref{fig:prop}.

In order to check the validity of this conjecture, we just consider the results obtained at $\beta$=5.6 as non-deviated and estimate the 
deviation parameter $\delta$ at $\beta$=5.8 required to get rid of the low-momentum discrepancies and get the data from both simulations lying on top of each other. This can be strikingly seen in the lower panel of Fig.~\ref{fig:prop}, to be left with which one needs to apply $\delta=-0.05$. Properly interpreted, the latter means that all the discrepancies can be explained if we accept a 5 \% of deviation in the ratio between the lattice spacings at $\beta$=5.8 and at $\beta$=5.6, with respect to the values quoted in Tab.~\ref{tab:setup}.  
These values have been obtained in~\cite{Guagnelli:1998ud} by using the Sommer parameter, $r_0$, and are compatible with those used in~\cite{Duarte:2016iko} and set by the string tension in~\cite{Bali:1992ru}. In both cases, the scale setting procedures refer to the force between external static charges. The relative accuracy of $r_0/a$ resulting from the thorough statistical analysis of~\cite{Guagnelli:1998ud} is of the order 0.3-0.6 \%, but a cut-off-dependent systematical uncertainty of 2-3 \% can be sensibly conceived and might be enough to explain the lattice spacing effects at low-momentum shown here and previously reported in~\cite{Duarte:2016iko}. Other scale setting prescriptions as the more precise one grounded on the Wilson flow~\cite{Luscher:2010iy, Luscher:2013vga, BMW} could presumably result on reduced systematic uncertainties. The comparison of the running of renormalized propagators can anyhow be of much help to check these uncertainties and refine the scale setting. 

\section{conclusions}
\vspace*{0.1cm}
We suggest that the lattice spacing effects discussed by the authors of~\cite{Duarte:2016iko}, taking place in the low-momentum domain of the quenched gluon and ghost propagators,  can be better justified by invoking small systematic deviations in the lattice scale setting based on the definition of the force between external static charges. 

\section*{Acknowledgements} 
\vspace*{-0.1cm}
We thank the support of Spanish MINECO FPA2014-53631-C2-2-P  research project, SZ acknowledges support by the National Science Foundation (USA) under grant PHY-1516509 and by the Jefferson Science Associates, LLC under U.S. DOE Contract \#DE-AC05-06OR23177.  %Numerical computations have used resources of CINES, GENCI-IDRIS and of the IN2P3 computing facility in France as well as of L-CSC in Germany

%-------------------------------------------------------
%-------------------------------------------------------

%%%%%%%%%%%%%%%%%%%%%%%%%%%%%%%%%%%%%%%%%%%%%%%%%%%%%%%%%
%%                              BIBLIO
%%%%%%%%%%%%%%%%%%%%%%%%%%%%%%%%%%%%%%%%%%%%%%%%%%%%%%%%%

%\bibliography{total}

\end{document}